\documentclass[prl,aps, 10pt, twocolumn, floatfix,superscriptaddress]{revtex4-1}
\usepackage{graphicx}
\usepackage{amsmath}
\usepackage{fullpage}

\begin{document}

\title{Implementing random unitaries in an imperfect photonic network}

\author{Roel Burgwal}
\author{William R. Clements}
\affiliation{Clarendon Laboratory, Department of Physics, University of Oxford, Oxford OX1 3PU, United Kingdom.}
\author{Devin H. Smith}
\author{James C. Gates}
\affiliation{Optoelectronics Research Centre, University of Southampton, Southampton SO17 1BJ, UK.}
\author{W. Steven Kolthammer}
\author{Jelmer J. Renema}
\email{jelmer.renema@physics.ox.ac.uk}
\author{Ian A. Walmsley}
\affiliation{Clarendon Laboratory, Department of Physics, University of Oxford, Oxford OX1 3PU, United Kingdom.}

\begin{abstract}
We numerically investigate the implementation of Haar-random unitarity transformations and Fourier transformations in photonic devices consisting of beam splitters and phase shifters, which are used for integrated photonics implementations of boson sampling. The distribution of reflectivities required to implement an arbitrary unitary transformation is skewed towards low values, and this skew becomes stronger the larger the number of modes. A realistic implementation using Mach-Zehnder interferometers is incapable of doing this perfectly and thus has limited fidelity. We show that numerical optimisation and adding extra beam splitters to the network can help to restore fidelity.
\end{abstract}

\maketitle

	Multiport interferometers are a crucial technology for optical communication and information processing, both in classical and in quantum optics. Classical applications include mode (de)multiplexers for few-mode fibers \cite{Miller2013,Melati2016},  self-aligning coupling into fiber \cite{Miller2013a}, and spatial-mode and polarisation converters \cite{Miller2013b}. On-chip multiport interferometers, consisting of an array of reconfigurable beam splitters (BSs) and phase shifters (PSs), are well suited for manipulation of photonic quantum states in e.g. quantum teleportation \cite{Metcalf2014}, quantum key distribution \cite{Honjo2004} or photonic qubit gates \cite{Carolan2015}, due to their inherent phase stability, reconfigurability and ease of fabrication.
    
One particular quantum-optical task which multiport interferometers are well suited for is boson sampling \cite{Aaronson2013}. The boson sampling task consists of sampling from the output photon number distribution of a large interferometer, which is fed with single-photon inputs. Since the first demonstrations \cite{Carolan2015,Crespi2013,Spring2013,Tillmann2013,Broome2013}, many advances have been made, by devising alternative sampling schemes that are easier to implement \cite{Hamilton2016,Bentivegna2015} and by improving the efficiency of single-photon sources \cite{Wang}. A direct implementation of this task in quantum hardware outperforms simulations on a classical computer for a not unreasonable number of photons, making it a promising technique for an unambiguous demonstration of a quantum advantage. 

The hardness proof of boson sampling requires that the unitary matrix that governs the mode transformation $a^\dagger_{\mathrm{out}} = U a^\dagger_{\mathrm{in}}$ is randomly chosen according to the Haar measure, and that the number of modes is much larger than the number of input photons. This has created interest in implementing large random unitary matrices in multiport interferometers \cite{Russell2015}, and the accuracy with which this can be done. 

	In this work, we study the implementation of Haar random unitaries in planar multiport interferometers with realistic fabrication tolerances. We use a recently developed decomposition algorithm \cite{Clements2016}, that implements a unitary transformation in a square array of BS-PS pairs. It can be shown that this decomposition has superior loss tolerance to an older decomposition \cite{Reck1994}, which uses a triangular arrangement. 
    	\begin{figure}
		\includegraphics[]{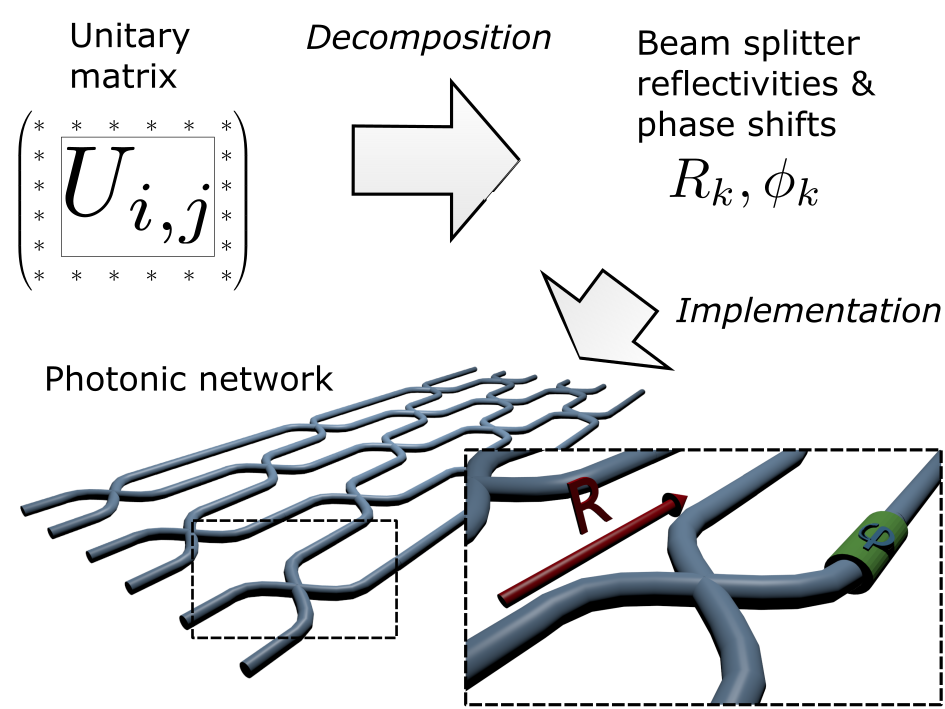}
        \caption{A unitary matrix can be implemented into a multiport interferometer via a mathematical decomposition. The interferometer consists of pairs of beam splitters and phase shifters (see inset). The square decomposition results in the structure of the interferometer shown.}
        \label{interferometer}
	\end{figure} 
    
    First, we find that an interferometer implementing random unitary matrices has interesting scaling properties. As the size of the interferometer increases, the majority of the beam splitters take on increasingly low reflectivities. Next, we find that for moderate interferometer sizes (20 modes) and realistic errors in fabrication, neither decompositions can implement any unitary transformation faithfully. Moreover, our results show that the allowable fabrication tolerances decrease with the size of the interferometer, meaning that any level of fabrication tolerance sets a limit on the size of a reconfigurable interferometer. 
    
    We also study techniques to mitigate this effect, by adding reduncancy to the system in the form of an additional layer of beam splitters and by numerically optimizing the beam splitter settings. We find that this for small networks, this technique increases the fidelity of the required tranformations to the $10^{-4}$ level, for realistic imperfections. This result points the way to the study of the robust creation of random matrices in photonic networks. 
    
    \section{Results}    
    
	Figure \ref{interferometer} shows the problem under study. We start with a unitary transformation we need to implement. The decomposition algorithms translate the unitary matrix into a set of beam splitter (BS) reflectivities ($R_k$) and phase shifts ($\phi_k$). These can then be implemented in a multiport interferometer, of which each node is a beam splitter and phase shifter pair (see inset).

    \begin{figure}[ht]
		\centering
		\includegraphics[]{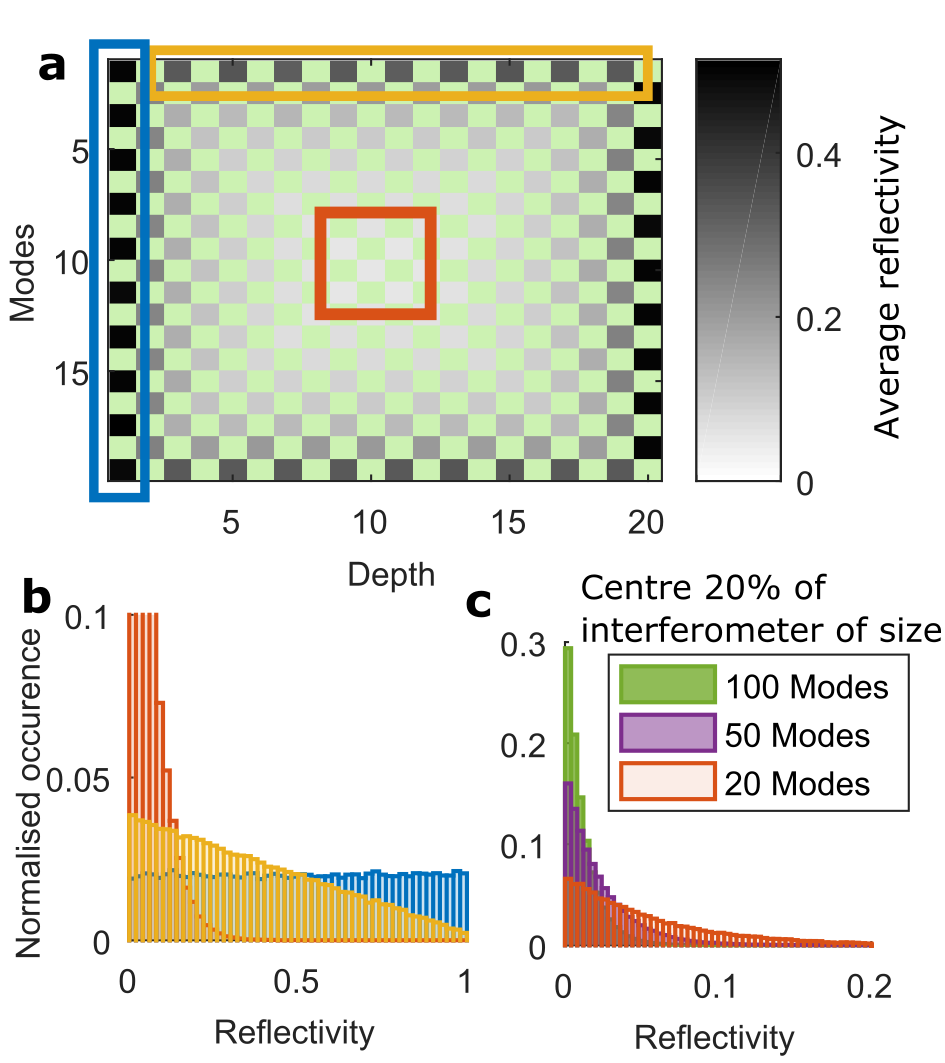}
        \caption{Interferometers implementing Haar-random unitary matrices show a specific distribution of beam splitter reflectivities. a) shows the spatial distribution of the average reflectivity in a size 20 interferometer. b) Shows the underlying histograms for three regions in the interferometer, the first column, top row and centre. c) shows how the centre-of-interferometer histogram scales with the size of the interferometer. Note the change of scale in c).}
        \label{reflectivities}
	\end{figure}
	
    Our first goal is to understand what the implementation of random unitary matrices looks like in terms of reflectivities and phase shifts. To do this, we performed the square decomposition on random unitary matrices. We calculated the average reflectivity for every BS in an interferometer of 20 modes, averaging over 5000 random unitary matrices. 
    
    Figure \ref{reflectivities}a shows the highly nonuniform spatial distribution of average reflectivity. Each grayscale square in the figure represents a beam splitter at the same location in the underlying interferometer, through which light travels from left to right. The modes are labeled along the \textit{y}-axes and the depth along the \textit{x}-axes. The colour indicates the average reflectivity, which ranges from 0 to 0.5. It is surprising that the centre of the interferometer has low values of reflectivity. In fact, the majority of beam splitters have low reflectivity and the overall average is 0.18. Note that low reflectivity means most light is transmitted, and thus travels along diagonal lines across the interferometer. Similar results can be found for the Reck decomposition by using the expressions for reflectivity distributions presented in \cite{Russell2015}.
    
	For Figure \ref{reflectivities}b, we have selected the three regions from the interferometer which are marked in subfigure a: the first column, top row and the interferometer centre, a square with sides of 20\% the interferometer size. For each of these we show the distribution of reflectivities that underlies the average of figure 2a, plotted in their corresponding colours. Most interesting is the distribution for the centre, which is peaked at low values and for which we found no values of the reflectivity larger than 0.4 in our sample. 
	
	Figure \ref{reflectivities}c shows that this effect becomes more pronounced as the size of the interferometer increases. The figure shows how the distribution of the centre of the interferometer changes with interferometer size. We have plotted the corresponding distribution for sizes 20, 50 and 100. The distribution becomes more sharply peaked at low values when increasing the size and the average reflectivity becomes lower. The distributions for the first column and top row do not change with size, thus the overall average reflectivity becomes lower as the interferometer size increases. From a similar analysis we found that the Reck decomposition also has this scaling property.
	
	These results can be understood intuitively through the properties of a Haar-random unitary matrix. Given a matrix $U$ that describes an interferometer, the amount of light that travels from input $j$ to output $i$ in a classical experiment is $|U_{i,j}|^2$. For Haar-random unitaries, the mean magnitude of every element is the same, $\langle |U_{i,j}|^2 \rangle = 1/N$. There is only one path, however, that light can take from input 1 to output $N$, on which light is transmitted at each BS (transmission $T = 1-R$): thus transmission has to be high and, correspondingly, reflectivity has to be low. 

	We now introduce the problem of interferometer imperfections. In particular, we investigate one type of imperfection that stands out when implementing Haar-random unitary matrices. Most reconfigurable realisations of multiport interferometers use Mach-Zehnder interferometers (MZIs) to implement variable beam splitters \cite{Carolan2015,Spring2013}. These interferometers contain two static 50:50 beam splitters. In practice, these beam splitters are not exactly 50:50, which means the MZI can generally not reflect or transmit all light. As shown above, low reflectivities are needed for the majority of MZIs in a large interferometer implementing random unitaries, thus this is problematic.

    \begin{figure}
\includegraphics[]{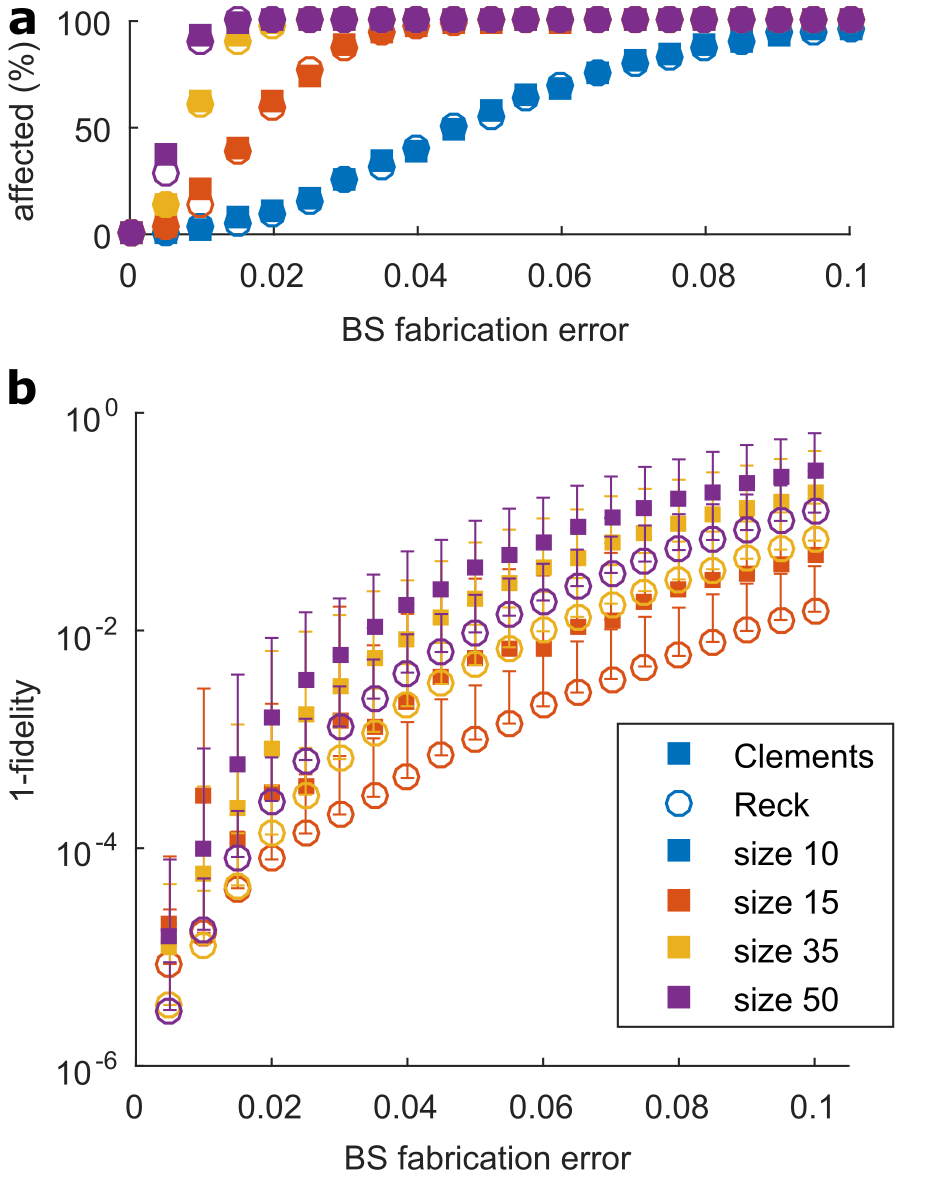}
\caption{The effect of unbalanced MZIs on the fidelity of the decompositions as a function of the size of the fabrication error. The triangular Reck decomposition and the square decomposition by Clements \textit{et al.} are used for various interferometer dimensions. a) us the fraction of the random unitaries that are affected by imperfections. b) the fidelity between the target and the effective unitary for the affected matrices when using our adapted version of the decompositions. The error bars show the standard deviation of all data points used in the average.}
\label{fidelity}
\end{figure}
	
    We quantified the error resulting from this limitation, using an adapted version of the decomposition. First, we generated a random unitary and decomposed it assuming a perfect interferometer. Next, we modeled the BS error: the reflectivities of the static BSs were drawn from a normal distribution with standard deviation $\sigma$ and mean 0.5. We refer to $\sigma$ as the fabrication error. Using these reflectivities, we calculated the minimum and maximum reflectivity of the corresponding MZI. We implemented the decomposition, setting reflectivities to the closest achievable value when they were out of reach. We then calculate the fidelity between the unitary achieved by this process and the target unitary as a measure of similarity.

Figure \ref{fidelity} shows us the effect of the imperfection when using this adapted decomposition. We have performed the adapted decomposition while varying the fabrication error of static BSs, which is displayed on the \textit{x}-axis. This we have done for various interferometer sizes up to size 50 and for both decompositions.

In figure \ref{fidelity}a, we show what fraction of the random unitaries is affected by the error. We see that, for larger interferometer sizes, unbalanced MZIs affect fidelity for even small fabrication error. This means that as MZI multiport interferometers grow in size, they are inevitably affected by the error at some point. The ratio is the same for both decompositions.
    
Figure \ref{fidelity}b shows the average fidelity for those random unitary matrices that cannot be implemented perfectly. The \textit{y}-axis shows one minus the fidelity, which means that a value of 0 implies the effective unitary is equal to the target. With current state-of-the-art fabrication tolerance (0.025,\cite{Mikkelsen2014,Kundys2009}), we are limited to 0.999 fidelity when building a 50-mode interferometer. To relate this value to experiment, we compare the results of single photon experiments of the effective matrix to the target unitary. We define $P^{\mathrm{exp}}$ as the set of single-photon transition probabilities of this implementation and $P$ as the same set for the target unitary. Then, for 0.999 fidelity, $\langle | P^{\mathrm{exp}}_i - P_i | \rangle / \langle P_i \rangle = 0.02$: probabilities are off by 2\% on average, with maximum averaging 25\%. The triangular interferometer is slightly more robust to these imperfections than the square interferometer. 

To compare the reflectivity distribution of random unitary matrices to those of other interesting interferometer applications, we have also performed the
new decomposition on the Fourier transform. The Fourier transform is used in several quantum algorithms, such as Shor’s algorithm \cite{Shor1997}, and is, like Boson Sampling, well-defined for any number of input modes. Moreover, it has the property that the transmitted  itensity is equal to the average transmitted intesity for a Haar-random matrix.

Similar to the decomposition of a Haar-random matrix, the resulting reflectivity distribution has low reflectivity on the diagonals and high values at the edges. However, low values of the reflectivity occur only exactly on the diagonal and not in the general center of the system. This means that the number of interferometers which is at risk of being affected by an imperfect beam splitter goes roughly as $N^2$ for a Haar-random unitary, but as $N$ for a Fourier matrix. 

This result has two implications. Firsts, it implies that Fourier matrices are in some sense 'easier' to implement than Haar-random matrices: they are affected by beamsplitter imperfections, but since there are fewer instances of low reflectivity, the probability of succesfully implementing a Fourier transformation on a system with given fabrication errors is higher than for a Haar-random unitary. This means that while Fourier matrices have been used as benchmarks for boson sampling \cite{Crespi2016}, they cannot be used as a benchmark the tunability of the interferometer. 

Second, it implies that our heuristic argument based on intensity transmission along the diagonals is incomplete: it holds for beam splitters which are exactly on the diagonals, but for those beam splitters which are slightly off-diagonal the phase relations must be considered as well.

\section{Numerical optimization and redundancy}

While both the triangular and square decompositions are unique, our strategy of dealing with imperfections in the interferometer locally is non-optimal. For cases where unity fidelity cannot be achieved, it is still possible to increase the fidelity by numerical optimization \cite{Mower2015,Spagnolo2016}.

Moreover, a natural strategy to mitigate the effect of imperfect components is to add redundancy to the network, in the form of additional beam splitters. Informally, these additional components provide paths with which to steer the light around the bottlenecks in the network. 
    
To study these ideas, we took the square  interferometer design and added an additional layer if beam splitters and phase shifters, respecting the quincunx layout. Similar to the techniques used to obtain figure 3, we then produce many random realistic networks and Haar-random matrices. We then numerically optimize our fidelity using the in-built sequential quadratic programming routine from MATLAB, subject to the constraints given by the limitations on our network. As our initial guess for the numerical optimization, we implement the square decomposition in the first $N$ by $N$ beamsplitters, and set the rest to be as reflective as possible given our constraints. We also performed direct numerical optimization on a square set of beam splitters, using the same techniques.

\begin{figure}
	\includegraphics[width=0.55\textwidth]{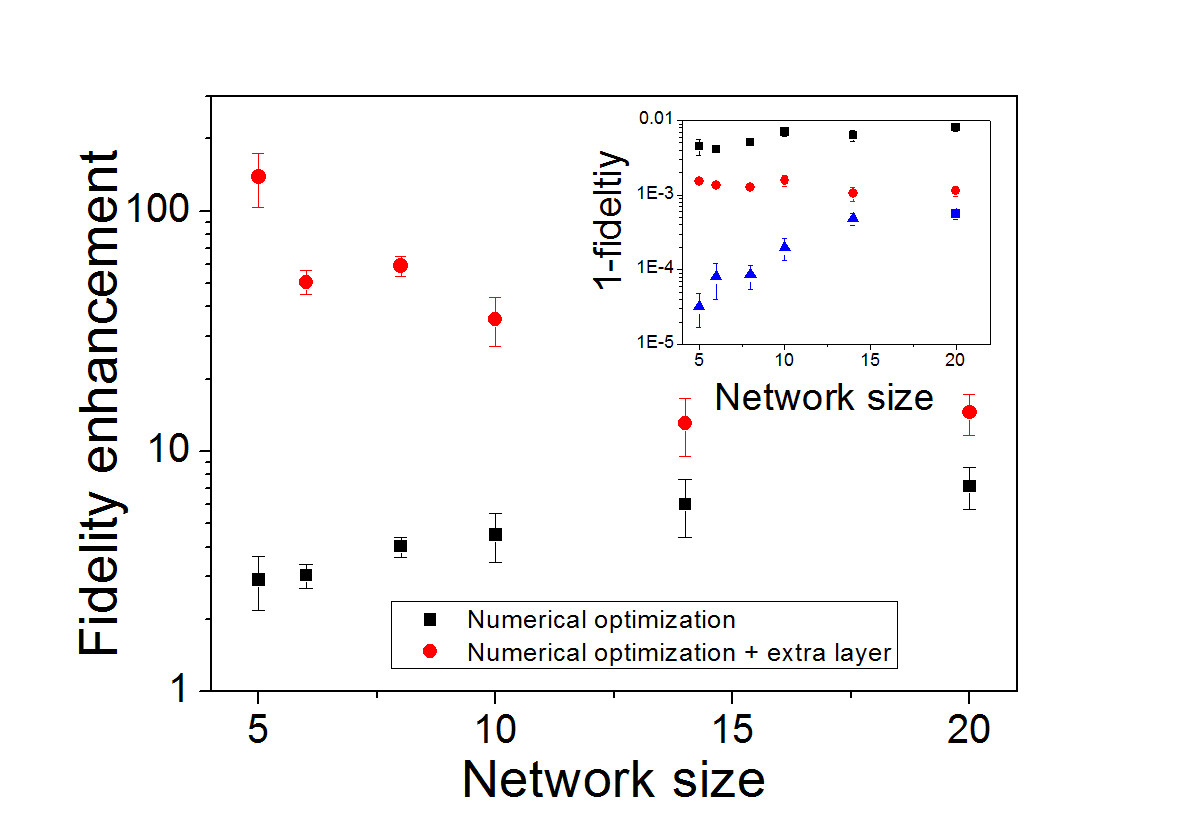}
    \caption{Numerical optimization and redundancy in a photonic network with realistic errors. The black points show the fidelity enhancement for numerical optimization only, the red points show enhancement when an additional layer of beam splitters is added. The inset shows the raw fidelities.}
    \label{optimisation}
	\end{figure}
    
Figure \ref{optimisation} show the enhancement in fidelity which is obtained by this procedure, given by the reduction of the infidelity, expressed as a ratio to the infidelity obtained with the direct approach. The black points show the enhancement given by numerical optimization, the red points show the enhancement given by adding an additional layer and then optimizing. The inset shows the raw fidelities. We sampled 100 combinations of matrices and sets of imperfections in each point.

For the case where we optimize the square network, we observe that the  fidelity enhancement increases with network size. As the network gets bigger, there is more scope for numerical optimization. In the extreme case of one beam splitter, there is no scope for optimization at all. 

Conversely, for the case where we add another layer, the enhancement decreases with network size. Again, this can be understood by considering the case of two modes. In that case, our scheme is identical to one independently proposed by Miller and Thomas-Peter \cite{Miller2015,NThomasPeter}, where the effect of unbalanced MZIs is largely circumvented by using two imperfect MZIs to implement one perfect variable BS. In this case, one expects complete fidelity. As we go to larger network sizes, the effect of a single additional layer becomes less noticable. We note that it is an interesting open problem how many additional layers are required as a function of network size to achieve constant fidelity, i.e. whether our redundancy approach is efficient or not.

Finally, we note that our optimization method is not efficient. The run-time of our numerical algorithm scales strongly with increasing network size. In the case of networks with inbuilt redundancy, the solver performs worse due to the fact that the unconstrained problem has more than one solution. We leave the question of an efficient algorithm to mitigate network imperfections as an open problem.

\section{Discussion}

Arkhipov showed \cite{Arkhipov2014} a bound on the overall distance between the target unitary and the unitary achieved by a network of optical components, given accuracy with which each component is set to its target value. However, this approach does not take into account the fact that not all points in the parameter space of component settings may be achieved equally easily in experiment, nor does it take into account the fact that the required settings may themselves be a function of network size, inducing further scaling behaviour. We have demonstrated that as the network grows, the required settings tend to move towards parts of the parameter space which - with the widely used implementation discussed here - are not easily accessible.
    
    
	In conclusion, we showed that the reflectivities in a multiport interferometer implementing Haar-random unitary matrices are such that fidelities are severely limited by unbalanced Mach-Zehnder Interferometers. We showed that, using optimisation of the parameters, some fidelity can be regained. More importantly, we found that slightly increasing the depth of the interferometer can improve the fidelity even in the presence of considerable error. This approach may also prove useful to mitigate the effects of other types of fabrication imperfections, such as unbalanced loss. The next step is to find a closed-form or low overhead method of finding these settings for a realistic interferometer with added depth. With such a solution in hand, one can greatly increase the fidelity of future large multiport interferometers.
    
    \section{Note}
    During the preparation of this manuscript, closely related results were published in \cite{newBristol}.
	
    \section{Acknowledgements}
    The  project  leading  to  this  application  has  received funding  from  the  European  Union's  Horizon  2020  research  and  innovation  programme  under  grant  agreement   No   665148. R.B. is supported by the Leiden University Fund (LUF) and the Minerva Scholarship Fund. J.J.R. is supported by the Netherlands Organization for Scientic Research (NWO). W.R.C. and I.A.W. acknowledge an ERC Advanced  Grant  (MOQUACINO). W.S.K. is supported by EPSRC EP/M013243/1. We thank P.G.R. Smith for discussions.

	\bibliographystyle{unsrt}

\begin{thebibliography}{10}

\bibitem{Miller2013}
D.~A.~B. Miller.
\newblock {Reconfigurable add-drop multiplexer for spatial modes}.
\newblock {\em Opt. Exp.}, 21(17):20220--20229, 2013.

\bibitem{Melati2016}
D.~Melati, A.~Alippi, and A.~Melloni.
\newblock {Reconfigurable photonic integrated mode (de)multiplexer for SDM
  fiber transmission}.
\newblock {\em Opt. Exp.}, 24(12):12625, 2016.

\bibitem{Miller2013a}
D.~A.~B. Miller.
\newblock {Self-aligning universal beam coupler.}
\newblock {\em Opt. Exp.}, 21(5):6360--70, 2013.

\bibitem{Miller2013b}
D.~A.~B. Miller.
\newblock {Self-configuring universal linear optical component}.
\newblock {\em Photonics Research}, 1(1):1, 2013.

\bibitem{Metcalf2014}
B.J. Metcalf, J~B Spring, P~C Humphreys, N~Thomas-Peter, M~Barbieri, W~S
  Kolthammer, X-M Jin, N~K Langford, Dmytro Kundys, J~C Gates, B~J Smith, P~G~R
  Smith, and I~A Walmsley.
\newblock {Quantum teleportation on a photonic chip}.
\newblock {\em Nat. Photon.}, 8(10):770--774, 2014.

\bibitem{Honjo2004}
T~Honjo, K~Inoue, and H~Takahashi.
\newblock {Differential-phase-shift quantum key distribution experiment with a
  planar light-wave circuit Mach-Zehnder interferometer.}
\newblock {\em Opt. Lett.}, 29(23):2797--2799, 2004.

\bibitem{Carolan2015}
J~Carolan, C~Harrold, C~Sparrow, E~Mart{\'{i}}n-L{\'{o}}pez, N~J. Russell, J~W.
  Silverstone, P~J. Shadbolt, N~Matsuda, M~Oguma, M~Itoh, G~D. Marshall, M~G.
  Thompson, J~C.~F. Matthews, T~Hashimoto, J~L. O'Brien, and A~Laing.
\newblock {Universal linear optics}.
\newblock {\em Science}, 349(6249):711--716, 2015.

\bibitem{Aaronson2013}
S~Aaronson and A~Arkhipov.
\newblock {The Computational Complexity of Linear Optics}.
\newblock {\em Theory Comput.}, 9(4):143--252, 2013.

\bibitem{Crespi2013}
A~Crespi, R~Osellame, R~Ramponi, D~J. Bord, E~F. Galv{\~{a}}o, N~Spagnolo,
  C~Vitelli, E~Maiorino, P~Mataloni, and F~Sciarrino.
\newblock {Integrated multimode interferometers with arbitrary designs for
  photonic boson sampling}.
\newblock {\em Nat. Photon.}, 7, 2013.

\bibitem{Spring2013}
J.~B Spring, B.~J. Metcalf, P.~C. Humphreys, W.~S. Kolthammer, X.-M. Jin,
  M.~Barbieri, A.~Datta, N.~Thomas-Peter, N.~K Langford, D.~Kundys, J.~C.
  Gates, B.~J. Smith, P~G~R Smith, and I~A Walmsley.
\newblock {Boson sampling on a photonic chip.}
\newblock {\em Science}, 339(6121):798--801, 2013.

\bibitem{Tillmann2013}
M~Tillmann, B~Daki{\'{c}}, R~Heilmann, S~Nolte, A~Szameit, and P~Walther.
\newblock {Experimental boson sampling}.
\newblock {\em Nat. Photon.}, 7(7):540--544, 2013.

\bibitem{Broome2013}
M~A. Broome, A~Fedrizzi, S~Rahimi-Keshari, J~Dove, S~Aaronson, T~C. Ralph, and
  A~G. White.
\newblock {Photonic Boson Sampling in a Tunable Circuit}.
\newblock {\em Science}, 339, 2013.

\bibitem{Hamilton2016}
C~S. Hamilton, R~Kruse, L~Sansoni, S~Barkhofen, C~Silberhorn, and I~Jex.
\newblock {Gaussian Boson Sampling}.
\newblock {\em arXiv:1612.01199}.

\bibitem{Bentivegna2015}
M.~Bentivegna, N~Spagnolo, C~Vitelli, F~Flamini, N~Viggianiello, L~Latmiral,
  P~Mataloni, D~J. Brod, E~F. Galv{\~{a}}o, A~Crespi, R~Ramponi, R~Osellame,
  and F~Sciarrino.
\newblock {Experimental scattershot boson sampling}.
\newblock {\em Sci. Adv.}, 1(3):e1400255, 2015.

\bibitem{Wang}
H~Wang, Y~He, Y~Li, Z~Su, B~Li, H~Huang, X~Ding, M~Chen, C~Liu, J~Qin, and
  J~Li.
\newblock {Multi-photon boson-sampling machines beating early classical
  computers}.
\newblock {\em arXiv:1612.06956}.

\bibitem{Russell2015}
N.~J. Russell, J.~L. O'Brien, and A.~Laing.
\newblock {Direct dialling of Haar random unitary matrices}.
\newblock {\em arXiv:1506.06220v1}.

\bibitem{Clements2016}
W.~R. Clements, P.~C. Humphreys, B.~J. Metcalf, W.~S. Kolthammer, and I.~A.
  Walmsley.
\newblock {An Optimal Design for Universal Multiport Interferometers}.
\newblock {\em Optica}, (2):8, 2016.

\bibitem{Reck1994}
M.~Reck, A.~Zeilinger, H~J. Bernstein, and P~Bertani.
\newblock {Experimental realization of any discrete unitary operator}.
\newblock {\em Phys. Rev. Lett.}, 73(1):58--61, 1994.

\bibitem{Mikkelsen2014}
J.~C. Mikkelsen, W.~D. Sacher, and J.~K.~S. Poon.
\newblock {Dimensional variation tolerant silicon-on-insulator directional
  couplers}.
\newblock {\em Opt. Exp.}, 22(3):3145--3150, 2014.

\bibitem{Kundys2009}
D.~O. Kundys, H~C. Gates, S~Dasgupta, C~B~E Gawith, and P~G~R Smith.
\newblock {Use of cross-couplers to decrease size of UV written photonic
  circuits}.
\newblock {\em IEEE Photon. Technol. Lett.}, 21(13):947--949, 2009.

\bibitem{Shor1997}
P.~Shor.
\newblock {Polynomial-Time Algorithms for Prime Factorization and Discrete
  Logarithms on a Quantum Computer}.
\newblock {\em SIAM Journal on Computing}, 26(5):1484--1509, 1997.

\bibitem{Crespi2016}
A~Crespi, R~Osellame, R~Ramponi, M~Bentivegna, F~Flamini, N~Spagnolo,
  N~Viggianiello, L~Innocenti, P~Mataloni, and F~Sciarrino.
\newblock {Suppression law of quantum states in a 3D photonic fast Fourier
  transform chip}.
\newblock {\em Nat. Commun.}, 7:10469, 2016.

\bibitem{Mower2015}
J~Mower, N~C. Harris, G~R. Steinbrecher, Y~Lahini, and D~Englund.
\newblock {High-fidelity quantum state evolution in imperfect photonic
  integrated circuits}.
\newblock {\em Phys. Rev. A}, 92(3):1--7, 2015.

\bibitem{Spagnolo2016}
N.~Spagnolo, E.~Maiorino, C.~Vitelli, M.~Bentivegna, A.~Crespi, R.~Ramponi,
  P.~Mataloni, R.~Osellame, and F.~Sciarrino.
\newblock {Learning an unknown transformation via a genetic approach}.
\newblock {\em arXiv:1610.03291}.

\bibitem{Miller2015}
D.~A.~B. Miller.
\newblock {Perfect optics with imperfect components}.
\newblock {\em Optica}, 2(8):747--750, 2015.

\bibitem{NThomasPeter}
N.~Thomas-Peter.
\newblock Quantum enhanced precision measurement and information processing
  with integrated photonics.
\newblock {\em PhD Thesis, University of Oxford}, 2012.

\bibitem{Arkhipov2014}
Alex Arkhipov.
\newblock {Boson Sampling is Robust to Small Errors in the Network Matrix}.
\newblock {\em Phys. Rev. A}, 92:062326, 2015.

\bibitem{newBristol}
N.~J. Russell, L.~Chakhmakhchyan, J.L. O'Brien, and A.~Laing.
\newblock Direct dialling of haar random unitary matrices.
\newblock {\em New J. Phys.}, 19:033007, 2017.

\end{thebibliography}

\end{document}